\documentclass[prd,superscriptaddress,nofootinbib,preprint]{revtex4}
 \usepackage{graphicx} \usepackage{dcolumn}
 \usepackage{bm}
 \usepackage{amsmath}
\usepackage{amsfonts}
\usepackage{amssymb}

 \usepackage{hhline} \usepackage[colorlinks]{hyperref}
 \usepackage[normalem]{ulem}

\begin{document} 
 \title{A simple model for magnetic inelastic dark matter (MiDM)} 
\author{Sudhanwa Patra} \email{sudha.astro@gmail.com} \affiliation{Physical Research
Laboratory, Ahmedabad 380009, India}
\affiliation{Institute of Physics, Bhubaneswar 751005, India
 } 
 \author{Soumya Rao} \email{soumya@prl.res.in}
 \affiliation{Physical Research Laboratory, Ahmedabad 380009, India}

 \begin{abstract} 

   A simple model for magnetic inelastic dark matter (MiDM), which is a minimal
   extension of the standard model with right-handed neutrinos ($N_R$), a singly
   charged scalar ($S$) and a vector-like charged fermion ($E$), has been presented.
   In this framework in which dark matter inelastically scatters off nuclei through
   a magnetic dipole interaction by making a transition to a nearly degenerate state
   with a mass splitting ($\delta$) of 100 keV.  The model is constrained from Direct
   detection experiments, DAMA and COGENT.  In our analysis we do not find any parameter
   space using the recent annual modulation data from COGENT, where DAMA and COGENT are
   consistent with each other. The bound from relic density provides a much stronger
   constraint.  We find that this minimal extension of SM incorporates a DM candidate
   which can explain the indirect detection results while being consistent with relic
   density measurements.  The right handed  neutrino and the charged particles   
   responsible for the right-handed neutrino magnetic moments could be produced at the
   Large Hadron Collider.  

 \end{abstract}

\maketitle

\section{Introduction} \label{sec:intro} 

The quest for the identification of dark matter (DM),together with the comprehension of
the nature of dark energy, is one of the most challenging problems in the understanding of
the physical world.  A canonical model for Dark Matter (DM) utilizing the existence of
Weakly Interacting Massive Particles (WIMPs) has emerged.  On the other hand,
supersymmetric extensions of the Standard Model naturally incorporate such stable
particles, like for example the neutralino.  The observed DM density, set by thermal
freeze-out, determines the cross-section to annihilate to Standard Model (SM) fields to be
a value typical of weak scale physics $\langle \sigma v \rangle \simeq 3.6\times 10^{-26}
\text{cm$^3$/sec}$. Within the paradigm of these models, many phenomenological
expectations have been fixed, including the annihilation modes to the SM interaction
channels with corresponding rates for indirect detection in the galaxy today.

Cosmological observations strongly suggest that non-luminous, non-baryonic matter
constitutes most of the dark matter in the universe. This dark matter should be
distributed in dark halos of galaxies such as the Milky Way, enabling the direct detection
of the dark matter particles via their interactions in terrestrial detectors. Recently,
inelastic dark matter (IDM) is a viable scenario to explain DAMA consistent with the other
experiments.  The inelastic scenario assumes that WIMPs ($\chi$) can only scatter off
baryonic matter (N) by transition into an excited state at a certain energy above the
ground state ($\chi N \rightarrow \chi^* N$), while elastic scattering is forbidden or
highly suppressed.

Since neutrinos have very weak interactions, any additional interaction, like the one
provided by magnetic moments, could have dramatic consequences in their behaviour. As a
result, neutrino magnetic moments provide new photon-neutrino couplings which affect the
production and detection of neutrinos at colliders and large transition magnetic moments
could affect neutrino oscillation. The magnetic moments of right-handed neutrinos arise as
dimension 5-operators which is the same dimension as of the well-known Weinberg operator
that is often used to parametrize neutrino masses. It is therefore natural to consider
right-handed neutrino magnetic moments as the first manifestation of non-trivial
electromagnetic properties of neutrinos. Among important low-energy properties of WIMPs
are their electromagnetic form factors. It has been known for a long time that the
possibility of charged WIMPs is strongly disfavored \cite{dm_emform_smith,gould} and
stringent limits exist in the case of a fractional charge \cite{dm_emform_Davidson}.

Motivated by the discussion above we present here a very simple model which gives rise to
right-handed neutrino magnetic moments; it includes, in addition to the SM fields and the
right-handed neutrinos, a charged scalar singlet and a charged singlet vector-like
fermion. While the phenomenological effects of an effective magnetic coupling of
right-handed neutrinos have their own intrinsic interests, their observation would also
indicate the presence of physics beyond the SM, and it is therefore natural to determine
the types of new particles and interactions that would be implied by the observation of
such effective couplings. There is, of course, a certain amount of freedom in constructing
models based on the single constraint that Majorana magnetic couplings be generated at the
one loop level, but the simplest possibility, involving the introduction of one charged
heavy scalar and one charged heavy fermion, involves the basic features of more general
models and addresses the main problems such theories face. In addition, if the
right-handed neutrino magnetic moments are large the new particles should be relatively
light and could be produced and analysed at the Large Hadron Collider (LHC) or the
International Linear Collider (ILC). This is also a general feature of models with large
right-handed neutrino magnetic moments.

Recently there has been a lot of excitement in the field of Dark Matter with the
observation of an annual modulation signal by COGENT \cite{Aalseth:2011wp}, similar to
what has been reported consistently by DAMA \cite{Bernabei:2008yi} over several years. An
excess in cosmic ray positron and electron signals over the expected background as
observed by PAMELA \cite{adriani_pamela} and FERMI \cite{abdo_fermi} may be a signal
of annihilating Dark matter for the energy of ${\cal O}$(100) GeV. The annihilation
cross-section needed to produce these signals is non-thermal, a factor of $\sim 10-1000$
(depending on DM mass and astrophysical boost factor) large than the thermal annihilation
cross-section \cite{weiner_boost,strumia_boost}. One possibility is to add new particles
and these particles mediate a Sommefeld enhancement, implying boosted annihilation in the
halo today, while also acting as intermediate final states, thereby allowing SM particles
produced from DM annihilations.

The paper is organised as follows. In section-\ref{sec:model}, we present a simple
model implementing inelastic dipolar dark matter which is a simple extension of the
Standard model for the dipolar dark matter to satisfy both obsevational and experimental
bounds. In the following section-\ref{sec:magmom}, we introduce the effective lagrangian
for the magnetic inelastic dark matter(MiDM) interaction with a photons. We discuss how
one can large magnetic dipole moment for heavy Majorana neutrino which is assumed to be
dark matter candiadate in our model and analytical expressions are presented in
section~\ref{sec:magmom}. We discuss the relic abundance in section~\ref{sec:relic}.
Section~\ref{sec:direct} presents constraints on dark matter diple moments and masses that
arises from direct search detectors DAMA and COGENT.  Comparison with the indirect
detection experiments PAMELA and FERMI is discussed in section~\ref{sec:indirect}.
Finally, we present our summary of our whole work in section~\ref{sec:summary}.

\section{Model} \label{sec:model} 

Beside the standard model particles and right-handed Majorana neutrinos, the model
contains a singly charged scalar and three vectorlike singly charged fermions. The new
fermions are assumed to be vector like to make sure that the theory is anomaly free as for
self consistency. The quantum numbers of these  extra particles under $SU(3)_c \times
SU(2)_L\times U(1)_Y$ gauge group : $N_R \equiv (1,1,0)$, $S \equiv (1,1,0)$ and $E \equiv
(1,1,0)$ and their $Z_2$ assignment is shown in Table:[\ref{tab:z2charge}].  The
hypercharge assignment is done by the relation $Y=Q-I_3$.

\begin{table}[!h] \begin{center} \caption{Particle content of the proposed Model }
  \label{tab:z2charge} \begin{tabular}{|c|c|c|c|} \hline &Field & $ SU(3)_C\times
    SU(2)_L\times U(1)_Y $ & $~Z_2~$ \\ \hline \hline Fermions&$Q_L \equiv(u,
    d)^T_L$        & (3, 2, 1/6)         & +  \\ &$u_R$
    & (3, 1, 2/3)         & +   \\ &$d_R$                          & (3, 1,
    -1/3)        & +    \\ &$\ell_L \equiv(\nu,~e)^T_L$    & (1, 2, -1/2)
    & +     \\ &$e_R$                          & (1, 1, -1)          & +
    \\ &$E$                            & (1, 1, -1)          & -      \\
       &$N_R$                          & (1, 1, 0)           & -       \\ \hline
       Scalars&$\Phi$                         & (1, 2, +1/2)        & + \\
       &$S^+$                          & (1, 1, +1)          & +    \\ \hline \hline
       \end{tabular} \end{center} \end{table}

We first analyze the case when the DM candidate is a fermion we denote by $\chi$, which is
assumed to be odd under $Z_2$ , and a gauge singlet; we will also assume that $\chi$ has
no chiral interactions. In this model, the Yukawa terms of the lepton sector would be

\begin{eqnarray} {\cal L}&=&-\bigg[Y_S~ \overline{(N_R)^c} E S^{+}+Y'_S~ \overline{N_R} E
  S^{+}+h.c.   \nonumber  \\ &-&\left[\frac{1}{2} \overline{(N_R)^C} M_N N_R-M_E
  \overline{E_L^c} E_L+h.c.\right] + V(\phi, \chi) 
\end{eqnarray} 
where $Y_S$ is the new Yukawa couplings, $M_R$ and $M_E$ are the mass matrices of $N_R$
and $E_L$.  In the unitary gauge the Higgs doublet is given by 

\begin{eqnarray} \nonumber
  \Phi=\frac{H+v}{\sqrt{2}} \begin{pmatrix}0\\ 1\end{pmatrix} 
\end{eqnarray} 
where $H$ is the SM Higgs boson and $v=246$ GeV is the vacuum expectation value.  All
through this study we will choose $M_H=120$ GeV.

The general scalar potential contains the quadratic and quartic terms as below

\begin{eqnarray} V(\Phi, S) &=& -m^2_1 \left( \Phi^\dagger\, \Phi \right) + \lambda_1\,
  \left( \Phi^\dagger\, \Phi \right)^2 +m^2_2 \left( S^\dagger\, S \right)
  +\lambda_2 \left( S^\dagger\, S \right)^2 \nonumber \\
  &+&\frac{\lambda_3}{2}\left( S^\dagger\, S \right) \left( \Phi^\dagger\, \Phi
  \right) 
\end{eqnarray}
where the mass of the singly charged scalar is $m^2_S=m^2_2+\lambda_2 v^2$.

 \section{Magnetic inelastic Dark Matter (MiDM)}\label{sec:magmom}

In this section, we focus on magnetic inelastic dark matter (MiDM), because it has a
unique and interesting directional signature and it has been shown that MiDM could explain
both DAMA and other null results \cite{weiner_dama}.  The model takes advantage of both
the magnetic moment and large mass of iodine.  In \cite{Fitzpatrick:2010br} the velocity
dependent scattering resulting from magnetic dipole operator is said to explain
simultaneously DAMA and COGENT results as well as null results from other experiments.
Explicit formula for scattering cross section assuming magnetic dipole operator has been
derived in \cite{Barger:2010gv}.

In MiDM, the dark matter couples off-diagonally to the photon:

\begin{equation}
\mathcal{L} \ni 
-\frac{\mu_\chi}{2}\, \overline{\chi^*}
\sigma_{\mu\nu}\, \mathcal{F}^{\mu\nu}\,\chi
\label{eqn:midm}
\end{equation}
where mass of $\chi$ and $\chi^*$ are split by $\delta=100$ keV. The off-diagonal coupling
is natural if the dark matter is a Majorana fermion. In the present scenario, heavy
Majorana neutrino is a candiadate for magnetic inelastic dark matter (MiDM). Due to the
Majorana nature, the magnetic moment of heavy Majorana neutrinos is zero. There is only
transition magnetic moment for them. The effective Lagrangian for coupling of a heavy
Majorana neutrino with a magnetic dipole moment $\mathcal{\mu}$ 

\begin{equation} 
\mathcal{L}_{edm}=-\frac{i}{2}\, \overline{N}_k
\sigma_{\mu\nu}(\mu_{jk})N_j 
\    \mathcal{F}^{\mu\nu}
\label{eqn:emdm_transition}
\end{equation}
where $\mu_{jk}$ is the transition magnetic dipole moment.

 \begin{figure}[ht] 
   \centering 
   \includegraphics[width=\textwidth]{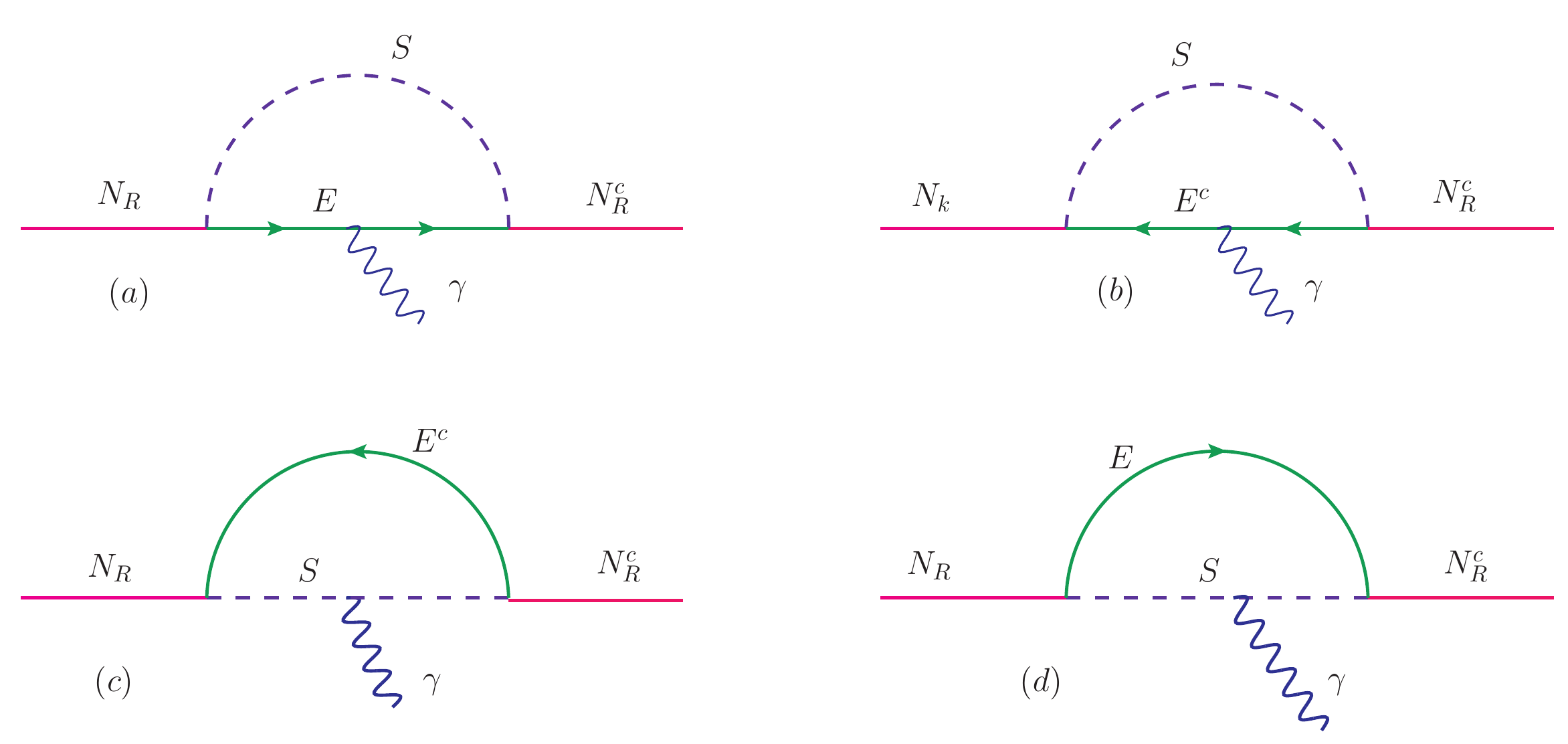}
   \caption{Contributing diagram for the heavy right handed Majorana neutrino}
   \label{fig:RHN_magnmom} 
 \end{figure} 
 
 \subsection{Magnetic moment of heavy RH Majorana neutrino} 

 In the model considered, we have four diagrams contributing to the transization magnetic
 moment, which are depicted in Fig.(\ref{fig:RHN_magnmom}): (a and b) a loop with the $B$
 gauge boson attached to the $E$ and (c and d) a loop with the $B$ gauge boson attached to
 the scalar $S$.

The Yukawa interactions of heavy Majorana neutrinos with $\chi$ and $E$ can be rewritten
in the following way 

\begin{equation} 
\frac{1}{2}\left[\overline{N^C_\alpha} S^-
(Y^T)_{\alpha i}^{} P_R^{} E_j^C + \overline{ N_\alpha^{}} S^+ (Y^\dagger)_{\alpha i}^{}
P_L^{} E_j^{}\right] + \frac{1}{2} \left[ \overline {E_j^{}} Y_{i\alpha}^{}
S^- P_R^{} N_\alpha^{}  + \overline{E_j^C} (Y^*)_{i \alpha}^{}  S^+ P_L^{}
N_\alpha^C\right] \; , 
\end{equation} 
through which we can derive relevant feynman rules.

Assuming that heavy Majorana neutrinos are nearly degenerate, i.e., $M_\alpha^{} \approx
M_\beta^{} \approx M_R$, we derive the expression of $\mu (0)$: 

\begin{equation}
  \mu_{\alpha \beta}^{}(0)= \frac{M_R}{64 \pi^2}\left [(Y_S^\dagger)_{\beta i}^{}
  (Y_S^{})_{i \alpha}^{} - (Y_S^T)_{\beta i}^{} (Y_S^{})_{i \alpha}^{*} \right]
  \left[ {\cal I} (M^2_S, M^2_R, M_{E_j}) -{\cal  I} (M_{E_j}, M^2_R, M^2_S)
  \right] \; , 
\end{equation} 
with 
\begin{equation} 
  {\cal I}(A, B, C) = \int d x \frac{x(1-x)^2}{(1-x) A + x(x-1) B + x C} \; ,
  \nonumber 
\end{equation}
where $M_{E_j}$ and $M_S$ are the mass eigenvalues of heavy vector-like  fermion $E$ and
singly charged scalar $S$, respectively.  In simplest form, one can write as 

\begin{equation} \label{eqn:magmom_RHN} 
  \mu_{\alpha \beta}^{}(0)\sim \frac{1}{32\pi^2} (Y_S^\dagger)_{\beta i}^{}
  (Y_S^{})_{i \alpha}\,\frac{e}{M_E} F(x)
\end{equation}
where $F(x)=\frac{1}{1-x}+\frac{x}{(1-x)^2} \ln(x), \quad x=\frac{M^2_S}{M^2_E}$.
Non-perturbatibe limit gives us  $(Y_S^\dagger\, Y_S) \leq 4 \pi$. With this spectrum, one
can get the magnetic moment of the order of $10^{-7} \mu_B$ to $10^{-9} \mu_B$. 

\subsection{Cross-section RH neutrino through \texorpdfstring{$e^+ e^- \to \gamma,\,
Z^*\to N_k N_j~(k \neq j)$}{e+ e- to photons, Z to RHN}]}

The most dramatic effect of a large magnetic dipole moment of a heavy neutrino will be in
the production cross section and angular distribution. Though the discussion of the
differential cross section for a heavy charged lepton can be found in work of Sher
\cite{Sher}, here we only discuss qualitatively how one can produce RH Majorana neutrinos
in near future experiment within our framework. In the discussion of Escribano and Masso
\cite{Masso}, one can write a $U(1)$ invariant operator as: $\overline{N}_{Rj}\,
(\mu^{jk}_N +i {\cal D}^{jk}_N)\, \sigma_{\alpha\beta}\, N^c_{Rk} \,B^{\alpha\beta}$,
where $B^{\alpha\beta}$ is the $U(1)$ field tensor. This gives a coupling to the photon,
which we define to be the EDM, as well as a coupling to the Z which is the EDM times $\tan
\theta_W$. When we include the effect of $Z$ coupling to $N$ in the differential cross
section, it turns out that the contribution has very little effect on the result.
 
The differential cross-section for the process, $e^+ e^- \to \gamma,\,Z^* \to N_k N_j~(k
\neq j)$, is given by
\begin{align}
\frac{d\sigma}{d \Omega}
             &
             =   \frac{\alpha^2}{4s} \sqrt{1-\frac{4M^2}{s}}\left( {\cal F}_1 
               + \frac{1}{8\sin^42\theta_W} P_{Z\, Z}\ {\cal F}_2 \right)
\nonumber \rule{0pt}{0.0em} \\ 
             &
               + \left(\frac{(1-4\sin^2\theta_W)\tan\theta_W }{\sin^22\theta_W} 
                  P_{\gamma\, Z}\ {\cal F}_3\right)
\rule{0pt}{2.0em}
\end{align}
where the values of ${\cal F}_1$, ${\cal F}_2$, ${\cal F}_3$, $P_{ZZ}$ and $P_{\gamma\,Z}$
\begin{align}
{\cal F}_1 &
           =  {\cal \mu}^2_N \, s \, \sin^2\theta \left(1+\frac{4 M^2}{s} \right)\;,
\nonumber \rule{0pt}{0.0em} \\ 
{\cal F}_2 &
           =  1+\cos^2\theta -\frac{4 M^2}{s}\sin^2\theta + 8 C_V \cos \theta
\nonumber \rule{0pt}{1.0em} \\ 
           & 
             +{\cal \mu}^2_N \, s \, \tan^2 \theta_W \left[\sin^2\theta 
             +\frac{4M^2}{s} \left(1+\cos^2\theta \right) \right] \;,
\nonumber \rule{0pt}{1.0em} \\ 
{\cal F}_3 
           &
           =  4 {\cal \mu}^2_N\, s \, \left[\sin^2\theta+ \frac{4 M^2}{s} 
              \left(1+\cos^2\theta \right) \right] \;,
\nonumber \rule{0pt}{1.0em} \\ 
P_{ZZ}    
           &
           =  \frac{s^2}{(s-M^2_Z)^2+\Gamma^2M^2_Z} \;,
\nonumber \rule{0pt}{1.0em} \\ 
P_{\gamma Z} 
          &
          =  \frac{s(s-M^2_Z)}{(s-M^2_Z)^2+\Gamma^2M^2_Z}.
\rule{0pt}{1.0em}
\end{align}
with $\mu^{kj}_N$ is the transition magnetic moment of heavy Majorana neutrino,
$C_V=\frac{1}{2}-2\sin^2\theta_W$, and we have dropped the numerically negligible
$C_V^2$
terms, for simplicity.
\begin{figure}[htb]
 \centering
 \includegraphics[width=10cm]{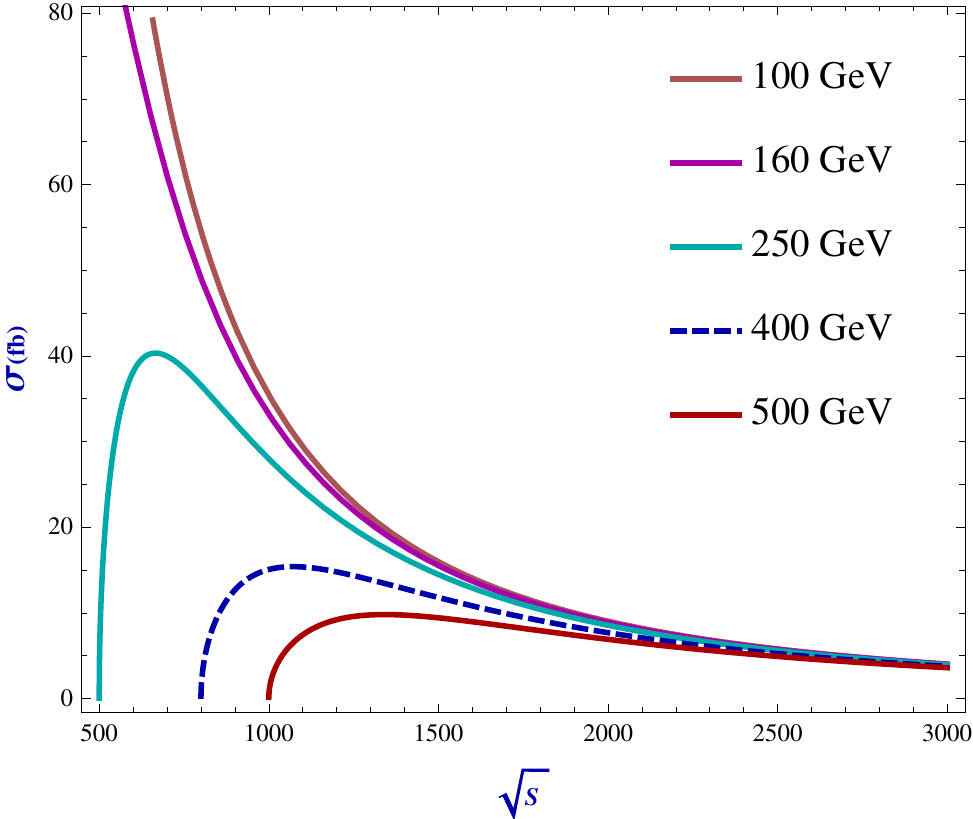}
\caption{The total cross section for heavy right handed neutrino 
$e^+ e^- \to \gamma,\,Z^* \to N_k N_j~(k \neq j)$ for various EDMs, 
in units of Bohr magneton. The cross section is shown as a function 
of center of collider energy $\sqrt{s}$ and here we have varied the 
masses of heavy right handed neutrino as $M=100, 160,250,400, 500$ GeV 
from the top to the bottom curves.}
\label{fig:prod_RHN_angular}
\end{figure}
\begin{figure}[htb]
 \centering
 \includegraphics[width=9cm]{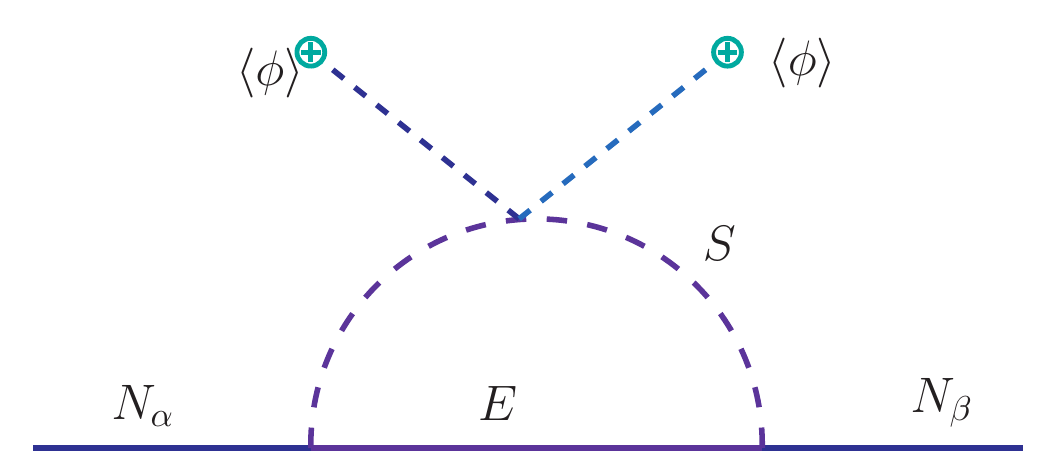}
 \caption{Mass splitting between the dipolar dark matter(DDM) and its companion partner}
 \label{fig:mass_splitting}
\end{figure}

\subsection{Mass splitting between two Majorana neutrino component}

In this paper we shall not discuss the origin of the small mass differences between the
degenerate right-handed neutrinos, but for completeness we demonstrate that a mass
splitting of the order of $10^{-7}$ GeV is not completely unnatural for 100 GeV scale
right-handed neutrinos.  Consider a diagram with a vertex $\lambda_{3} (\Phi^\dagger \Phi)
(S^\dagger S)$ attached to the singly charged scalar $S$ which runs in loop and this kind
of diagram gives a finite contribution to the mass splitting. The diagram with a vertex
$\lambda_{3} (\Phi^\dagger \Phi) (S^\dagger S)$ attached to the $S$ field (which is shown
in fig:(\ref{fig:mass_splitting}) give a finite contribution to the mass splitting as:  
\begin{align}
{\Delta M}_R  &
               \sim \frac{\lambda_{3} Y^*_S Y_S}{(4 \pi)^2} \frac{\langle \Phi \rangle^2}{4 M_E}
\end{align}
For the mass of the charged lepton to be around 100 GeV (i.e, $M_E\sim $ 100-150 GeV),
$\langle \Phi \rangle$=100 GeV and $Y^2_S \sim 1-10$, one can write
\begin{align}
{\Delta M}_R &
      \sim \frac{10^{5}}{64 \pi^2 M_E} \lambda_{S \Phi}
\end{align}
Now one can easily get the mass splitting between two right handed neutrinos of the order
of {\cal O}($10^{-4}$) GeV. If we thus start with a symmetry to get a 100 GeV Scale
degenerate right-handed neutrinos, after the symmetry breaking, we get a mass splitting
between the companion states of right-handed neutrinos to be in the range of {\cal
O}($10^{-4}$) GeV or 100 keV range, naturally through radiative corrections.

\section{Dark Matter annihilation and Relic Abundance}\label{sec:relic}

To calculate the relic abundance we solve the Boltzmann equation for heavy Majorana
neutrino which is assumed to be a Dipolar Dark Matter candidate, given by \cite{boltz}  

\begin{equation}
\label{eqn:relic-a} \nonumber
\frac{d\,n_{\chi}}{dt}+3\,H\,n_{\chi}=-\langle \sigma \,v \rangle
\left[n_{\chi}^{2}-n_{\chi, \rm eq}^{2} \right]
\end{equation}\noindent
where $H$ is the Hubble parameter, $n_{\chi}$ is the number density of Dipolar Dark Matter
($\chi$), $n_{\chi, \rm eq}^{2}$ is the equilibrium number density, $v$ is the relative
velocity and $\langle \sigma \,v \rangle$ is the thermal average of the annihilation
cross-section $\chi \chi \rightarrow \text{all}$.

Since the Dipolar Dark Matter considered in our model is non-relativistic, the equilibrium
number density is $n_{\chi, \rm eq}= g_{\chi} (M_{\chi}^2/(2 \pi x))^{3/2}\, e^{-x}$
produced by the back-reaction $f f^\prime \rightarrow {\chi}~ {\chi}$  at a given
temperature, where $g_{\chi}$ is the spin degrees of freedom, $T$ is the temperature and
$M_{\chi}$ is the mass of the relic. For particles which may potentially play the role of
cold dark matter, the relevant temperature is order of $\frac{M_{\chi}}{20}$ and the
non-relavistic equilibrium abundance is well justified.

In terms of the dimensionless variables $Y_{\chi}\equiv n_{\chi}/s$ where $s$ is the
total entropy density, and $x\equiv \frac{M_{\chi}}{T}$ the boltzmann equation becomes

\begin{equation}
\label{eqn:relic-dydx}
\frac{d Y_{\chi}}{dx}= - \frac{\lambda(x)}{x^2} \left[ Y^2_{\chi}-Y^2_{\chi,\rm eq}
\right]
\end{equation}
where 
\begin{eqnarray}
& & \lambda(x) \equiv \left(\frac{\pi}{45}\right)^{1/2}\,M_{\rm pl} M_{\chi}
\left(\frac{g_{*s}}{\sqrt{g_{*}}} \right) \, \langle \sigma v \rangle(x) \\
& &Y_{\chi, \rm eq}=\frac{45}{2 \pi^4} \left(\frac{\pi}{8}\right)^{1/2}
\frac{g_{\chi}}{g_{*s}} x^{3/2} e^{-x}
\end{eqnarray} 
where $g_{*}$ and $g_{*s}$ are the effective degrees of freedom of the energy density and
entropy density respectively. Here $M_{\rm pl}=2.4 \times 10^{19}$ GeV is the Planck mass
and $M_{\chi}$ is the mass of Dipolar Dark Matter ($\chi$).

\begin{figure}[htb]
 \centering
 \includegraphics[width=8cm]{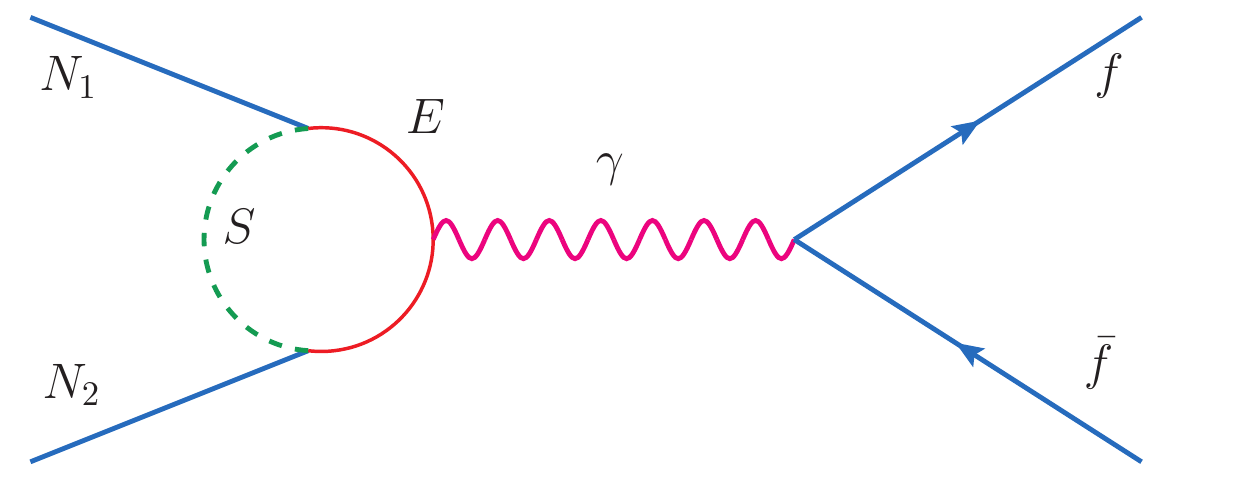}
 \caption{Annihilation diagram for dipolar dark matter(DDM) pair into fermion
 particle-antiparticle pair}
 \label{fig:ann_DDM}
\end{figure}

In the case of Majorana fermions only non-identical fermions can annihilate ($\chi_1 \bar
\chi_2 \rightarrow f \bar{f}$) through the dipole moment operator as only the transition
dipole moments ($\mathcal{D}_{12}, \mu_{12}$) are non-zero.  When the mass difference
$\delta$ between $\chi_1$ and $\chi_2$ is small, the cross section for Majorana
annihilation process is identical to that of the Dirac fermions (with $ {\cal D}, \mu$
replaced with ${\cal D}_{12}, \mu_{12}$).  The iDDM pairs annihilate to either photons or
charged pairs through the diagrams shown in Fig:
  
Hence the total cross-section is $\sigma_{\rm ann}\, v= \sigma_{\chi \chi^* \rightarrow 2
\gamma}\, v + \sigma_{f \bar{f} \rightarrow 2 \gamma}\, v$. Here we will calculate the
comsological relic abundance $\Omega h^2$ of the iDDM by asuuming standard freeze-out of
annihilations via the dipole coupling to $\gamma$ and there is no $\chi \overline{\chi}$
asymmetry. In the nonrelativistic limit, the thermal average annihilation cross section
$\langle \sigma\, v \rangle$ reduces to an average over a Maxwell-Boltzmann distribution
function \cite{boltz}: 

\begin{equation}
\label{eqn:sigma}
\langle \sigma\, v \rangle = \frac{x^{3/2}}{2 \pi^{1/2}} \int^1_0  
(\sigma \, v)\, v^2 \, e^{-\frac{x v^2}{4}} dv 
\end{equation}
The DM annihilation rate for magnetic dipolar interaction, with fermions as the
annihilation products is given by \cite{Masso:2009mu}

\begin{equation}
 (\sigma v)=\frac{e^2\mu_{12}^2}{4\pi}
\end{equation}

However we find that the dominant contribution to relic density comes from the
annihilation through magnitic dipole interaction into $W^+W^-$ given by

\begin{equation}
 (\sigma v)=\frac{3e^2\mu_{12}^2}{4\pi}\left(6-\frac{m_W^2}{m_\chi^2}\right)
\end{equation}



The present day mass density of iDDM particles $\chi$ is given by 
\begin{eqnarray}
\Omega_{\rm DDM,0} h^2 &\sim& \frac{x_f}{\sqrt{g_*}} 
           \frac{8.5\times 10^{-11}\, \text{GeV}^{-2}}{\langle \sigma\, v \rangle}
\end{eqnarray}
where we take $x_f=20$.

The observed value of CDM density from the seven year WMAP data  is $\Omega_{\rm CDM}\,
h^2=0.1123 \pm 0.0105$ (3$\sigma$) \cite{Komatsu} (where $h$ is the Hubble parameter in
units of $100$ km/s/Mpc). The annihilation cross-section required to get this value of
$\Omega h^2$ is $1.81\times 10^{-26} \text{cm$^3$s$^{-1}$}$.  To obtain this value of
annihilation cross section in our model 
we choose a DM mass of $160$ GeV which gives $\mu=3.8 \times 10^{-7} \mu_B$.



\section{Direct Detection of MIDM} \label{sec:direct} 

In direct detection of DM one measures WIMPs scattering off target nuclei.  The scattering
can be assumed to be either elastic or inelastic.  In the case of inelastic scattering,
the process is $\chi_1 +N\rightarrow \,\chi_2+N$ where $\chi_1$ and $\chi_2$ are two
different mass eigenstates, and in general, there is a mass difference between $\chi_1$
and $\chi_2$, $\delta=m_2-m_1$. Due to this mass difference, the minimum DM kinetic energy
needed for the nucleon scattering becomes higher \cite{weiner}. There is a minimal
velocity required to produce recoil energy $E_R$ in such an inelastic scattering

\begin{equation} \label{vmin} v_{\rm min}=\sqrt{\frac{1}{2 m_N E_{R}}}\left(\frac{m_N
  E_{R}}{\mu}+\delta\right) 
\end{equation} 
where $m_N$ is the mass of the target nucleus, $\mu$ is the reduced mass of the
WIMP-nucleus system, and $\delta$ is the WIMP-mass splitting.  Here we assume the WIMPs to
have a Maxwellian velocity distribution as described in \cite{dm_susy_griest}.


\begin{figure}[htb] \centering \includegraphics[width=8cm]{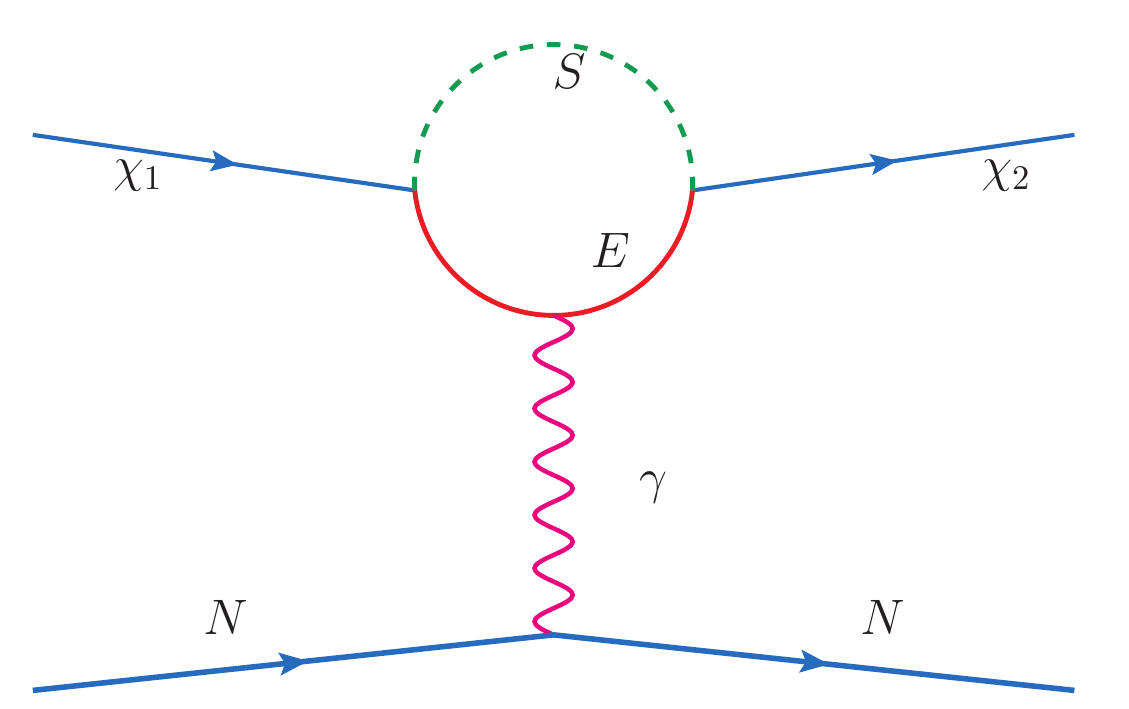}
 \caption{Inelastic scattering of MiDM particle with a nucleus through the magnetic moment
 of fermion singlet induced by a charged lepton loop and a photon vertex at 1-loop}
\end{figure}
\begin{figure}[ht] 
  \begin{center}
    \includegraphics{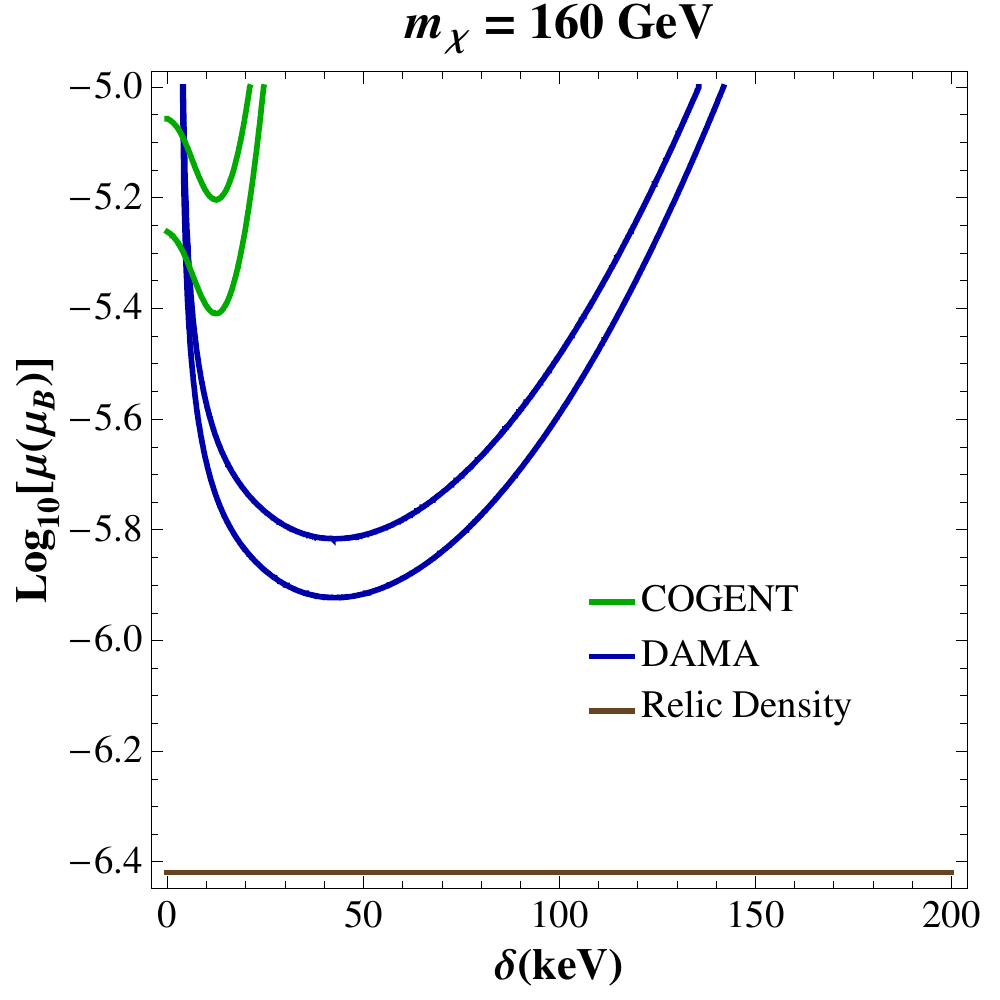}
  \end{center}
  \caption{Allowed parameter space in the $M_\chi$-$\delta$ plane for COGENT and
  DAMA.  Solid line represents bound from relic density.}
  \label{fig:direct}
\end{figure}

The differential cross section per unit energy transfer for elastic scattering by a
magnetic dipole moment interaction is given by \cite{Barger:2010gv}

\begin{equation} \frac{d\sigma}{dE_R}
={e^2 \mu^2_\chi\over 4\pi E_R}
\left[ Z^2 \left(1-{ E_R\over 2 m_N  v^2}-{ E_R\over m_\chi v^2}
\right) 
+ {I+1\over 3I}\left( {\mu_N\over {e\over 2m_p}} \right)^2 
{m_N E_R\over  m_p^2 v^2}\right] |F(q)|^2
\label{dsde}
\end{equation} 
where the first term is due to interaction with the charge of the nucleus and the second
term with the magnetic dipole moment of the nucleus.  Here $I$ denotes the nuclear spin,
$\mu_N$ and $\mu_\chi$ denote the magnetic moments of nucleus and DM respectively, Z
is atomic number of the target nucleus and $|F(q)|^2$ is the nuclear form factor. 
Although the nuclear charge form factor and the nuclear magnetic moment form factor can
be different we assume them to be approximately equal for simplicity
\cite{Barger:2010gv}.  The differential cross section given by \ref{dsde} is used to
calculate the event rate spectrum.  We use the recently released modulation data from the
COGENT experiment \cite{Aalseth:2011wp} as quoted in \cite{Frandsen:2011ts} and also the
latest modulation data from DAMA \cite{Bernabei:2008yi}.  The allowed parameter space
extracted assuming magnetic dipolar interaction is shown in Fig.~\ref{fig:direct}. It can
be seen from the figure that DAMA and COGENT can both be explained for a small sliver of
parameter space for a very small mass difference of about 4-5 keV. Also the bound
from relic density is a much more stringent constraint compared to the direct detection
bound from DAMA or COGENT.

\section{Indirect Detection of MIDM} \label{sec:indirect} 

The recently reported excesses of positrons and antiprotons in cosmic rays by the
experiments like PAMELA \cite{adriani_pamela,Pamela-pbar1,Pamela-pbar2} and of electrons+
positrons by Fermi\cite{abdo_fermi,FermiLat-electron1,FermiLat-electron2} can be explained
by dark matter annihilation with possibly a boost factor required in the DM annihilation
cross section.  This boost factor is explained either by astrophysical sources or effects
like Sommerfeld enhancement.

We fix the DM mass to be 160 GeV which gives the correct relic density for a DM
annihilation cross section of $1.81\times 10^{-26}{\rm cm}^3{\rm s}^{-1}$.  The dominant
contribution to relic density comes from the $W$ channel as shown in the previous section.
 But in order to obtain the positron and antiproton fluxes for the indirect detection
analysis, we take contributions from $W$ as well as the leptonic channels $e,\,\mu ,\,$
and $\tau$.  In case of $e$ the cross section will be a $\delta$-function.

\begin{figure}[ht]
  \begin{center}
    \includegraphics{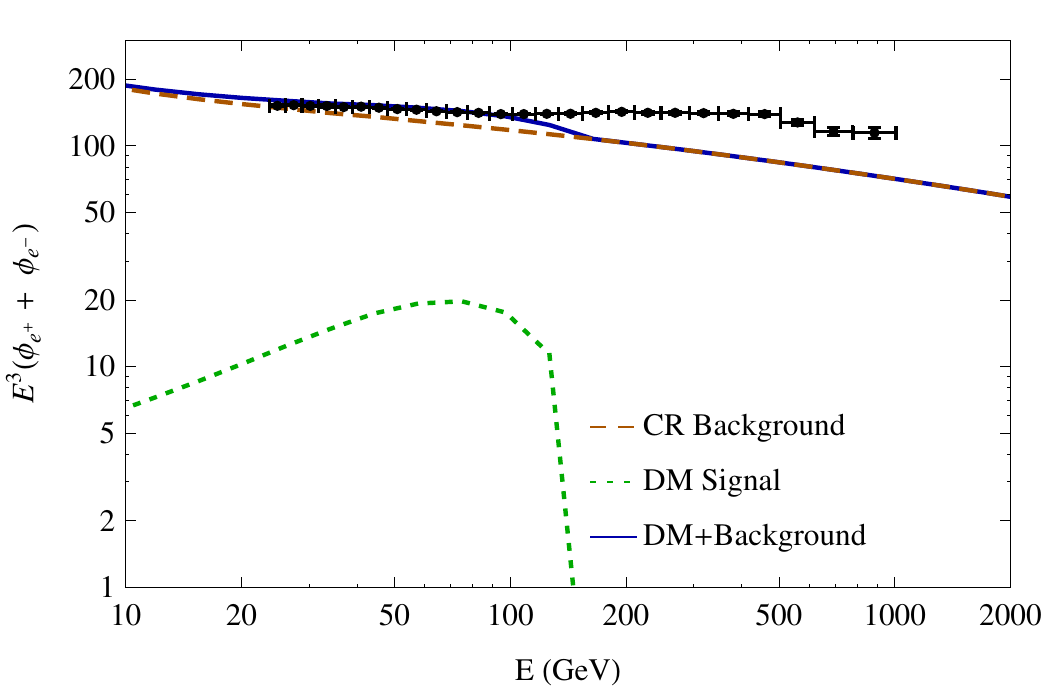}
  \end{center}
  \caption{The $(e^- + e^+) $ flux  for the 160 GeV DM compared with
  FERMI-LAT data \cite{FermiLat-electron1, FermiLat-electron2}. Dashed denotes the
  CR background and dotted line is the DM annihilation signal.}
  \label{fig:fer}
\end{figure}

\begin{figure}[ht]
  \begin{center}
    \includegraphics{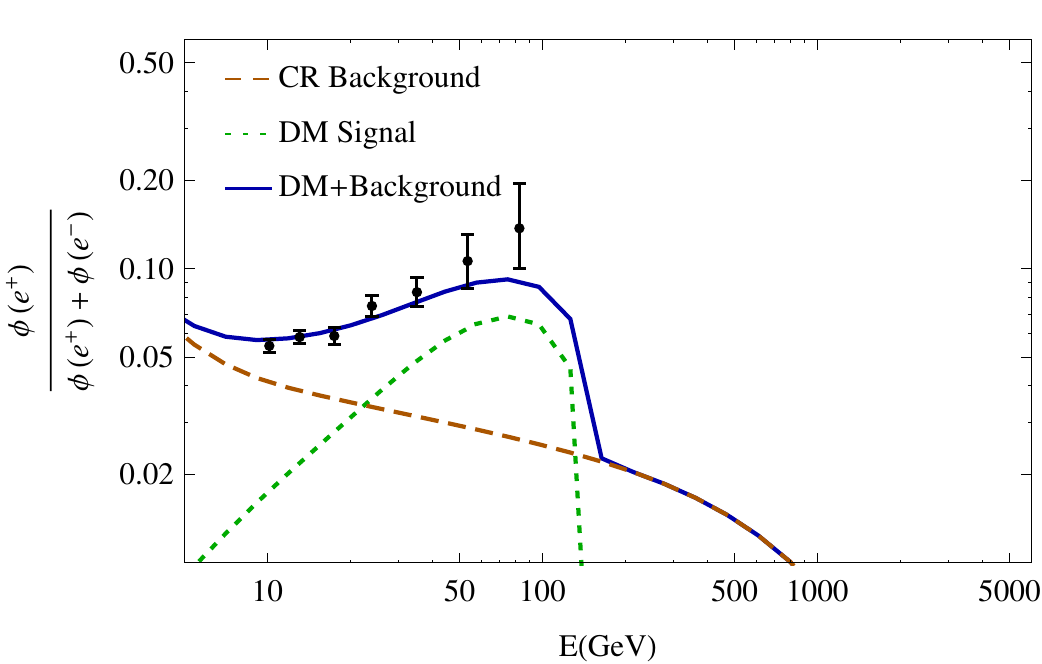}
  \end{center}
  \caption{Positron flux ratio for the 160 GeV DM compared with  Pamela data
  \cite{adriani_pamela}.  The annihilation cross section is taken to be $\sigma
  v_{rel}= 1.81 \times 10^{-25} cm^3 s^{-1}$ which corresponds to a boost of $10$.}
  \label{fig:pam}
\end{figure}

\begin{figure}[ht]
  \begin{center}
    \includegraphics{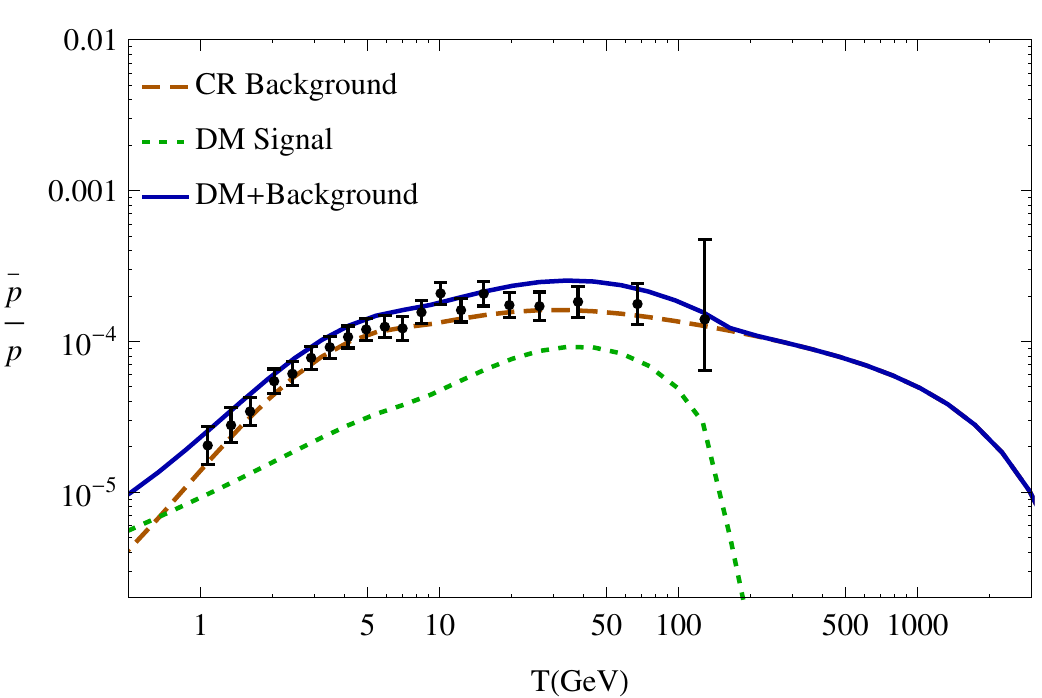}
  \end{center}
  \caption{Antiproton/Proton flux ratio  for the 160 GeV DM compared with Pamela data
  \cite{Pamela-pbar1,Pamela-pbar2}. The annihilation cross section is taken to be
  $\sigma v_{rel}= 1.81\times 10^{-25} cm^3 s^{-1}$ for all annihilation channels with a
  boost of $10$.}
  \label{fig:pbar}
\end{figure}

The method followed here to obtain the positron and antiproton fluxes is outlined in
\cite{Mohanty:2010es}.  We use \cite{pppc} to obtain event spectra for positrons and
antiprotons produced from $l\bar l$ ($l=e,\,\mu,\,\tau$) and $W^+W^-$.  This is then fed
into GALPROP \cite{Moskalenko:1997gh,Strong:2007} which calculates the final fluxes of
positrons and antiprotons to be compared with experiment.  The diffusion parameters used
in the Galprop code are as tabled in \cite{Mohanty:2010es}.  The DM annihilation cross
section used for calculating the positron and antiproton fluxes is $1.81\times
10^{-25}{\rm cm}^3{\rm s}^{-1}$, a boost factor of $10$ is required to fit the data.  Our
results are shown in Fig.~\ref{fig:fer}, Fig.~\ref{fig:pam} and Fig.~\ref{fig:pbar}.  The
positron flux shows fairly good agreement with the PAMELA data \cite{adriani_pamela} as
seen in Fig.~\ref{fig:pam}.  The antiproton flux is also within the observed values as
shown in Fig.~\ref{fig:pbar} due to a relatively light DM mass of 160 GeV and a very small
boost in the cross section.

\section{Summary}\label{sec:summary}

We have consider a minimal extension of the standard model which can explain non-zero
light neutrino mass, magnetic moment of heavy RH singlet neutrino which we consider here
as a DM candidate.  We have shown that the inelastic dipole dark matter particle (which is
a heavy RH singlet fermion in this case) with non-zero electric and/ or magnetic dipole
moment can satisfy the experimental and observational bounds.  Furthermore, this scenario
may be tested at future particle colliders (such as the Large Hadron Collider (LHC) ) or
dark matter detection experiments.

At the same time, it gives the connection between neutrino mass and magnetic dipole
coupling which has opened up new possibilities for model builders. We have considered the
dipole moment interactions between the heavy-light and heavy-heavy neutrino counterparts.
As a result, the large magnetic moments of heavy Majorana neutrinos can enhance the
production cross section of TeV scale right-handed neutrinos though the Drell-Yan process,
$e^+ e^- \to \gamma, Z^* \to N_i N_j~(i \neq j)$, which is within the reach of the future
linear collider (ILC).

The model also provides a good agreement with the indirect detection experiments like
PAMELA and FERMI consistent with the relic density bound from WMAP.  At the same
time we also constrain magnetic dipole moment from Direct detection experiments like DAMA
and COGENT. In our model we do not find a common parameter space for DAMA and COGENT
using the annual modulation data from both experiments. We also find that the relic
density constraint is much more stringent compared to constraints from DAMA and COGENT.


\end{document}